\documentclass[10pt]{emulateapj}
\usepackage{graphicx}
\usepackage{amsmath}
\usepackage{apjfonts}
\usepackage{natbib}

\begin{document}

\title{Dust Scattering and the Radiation Pressure Force in the M82 Superwind}

\author{Carl T.~Coker, Todd A.~Thompson, \& Paul Martini}

\affil{Department of Astronomy and Center for Cosmology \& Astro-Particle Physics,
The Ohio State University, Columbus, Ohio 43210, USA\\
coker, thompson, martini@astronomy.ohio-state.edu}

\begin{abstract}

Radiation pressure on dust grains may be an important physical mechanism driving galaxy-wide superwinds in rapidly star-forming galaxies. We calculate the combined dust and gas Eddington ratio ($\Gamma$) for the archetypal superwind of M82. By combining archival GALEX data, a standard dust model, Monte Carlo dust scattering calculations, and the Herschel map of the dust surface density distribution, the observed FUV/NUV surface brightness in the outflow constrains both the total UV luminosity escaping from the starburst along its minor axis ($L_{\rm\star,\, UV}$) and the flux-mean opacity, thus allowing a calculation of $\Gamma$. We find that $L_{\rm\star,\, UV}\approx 1-6\times 10^{42}$\,ergs s$^{-1}$, $\sim2-12$ times greater than the UV luminosity observed from our line of sight.  On a scale of $1-3$\,kpc above the plane of M82, we find that $\Gamma\sim 0.01 - 0.06$.  On smaller scales ($\sim0.25-0.5$\,kpc), where the enclosed mass decreases, our calculation of $L_{\rm\star,\, UV}$ implies that $\Gamma\sim0.1$ with factor of few uncertainties.  Within the starburst itself, we estimate the single-scattering Eddington ratio to be of order unity.  Thus, although radiation pressure is weak compared to gravity on kpc scales above the plane of M82, it may yet be important in launching the observed outflow.  We discuss the primary uncertainties in our calculation, the sensitivity of $\Gamma$ to the dust grain size distribution, and the time evolution of the wind following M82's recent starburst episodes.

\end{abstract}


\section{Introduction} \label{sec:introduction}

Galactic winds are crucial to the evolution of galaxies (see, e.g., Heckman et al.\ 1990).  Several mechanisms have been proposed for driving them, including heating from supernovae (Chevalier \& Clegg\ 1985) and stellar winds, heating from photoionization (Shapiro et al.\ 2004), and radiation pressure on dust grains (Murray et al. 2005, 2011, Zhang \& Thompson\ 2012). Recent simulations indicate that radiation pressure may indeed be an important driver in galaxies with high star formation rates and gas densities (Hopkins et al.\ 2012).  In the radiation pressure picture, radiation from hot stars drives dusty gas out of individual star-forming clouds, disrupting them, cutting off star formation, and then lofting this gas above the plane, where radiation from the entire galactic disk drives the gas fully out of the galaxy as a galactic superwind (Murray et al.\ 2011, Hopkins et al.\ 2012).

\begin{figure*}[t]
\centerline{
\includegraphics[scale=0.4]{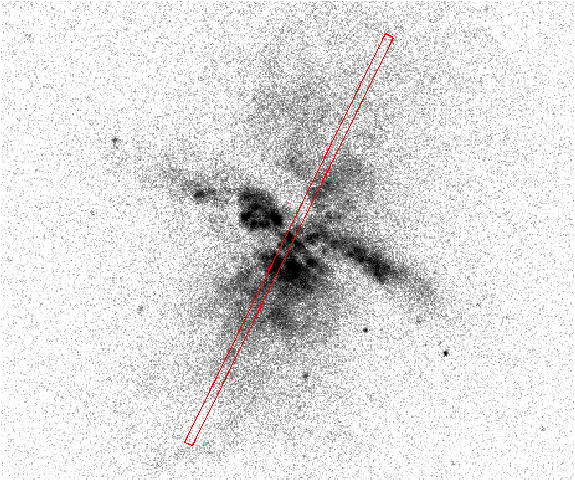}
\includegraphics[scale=0.4]{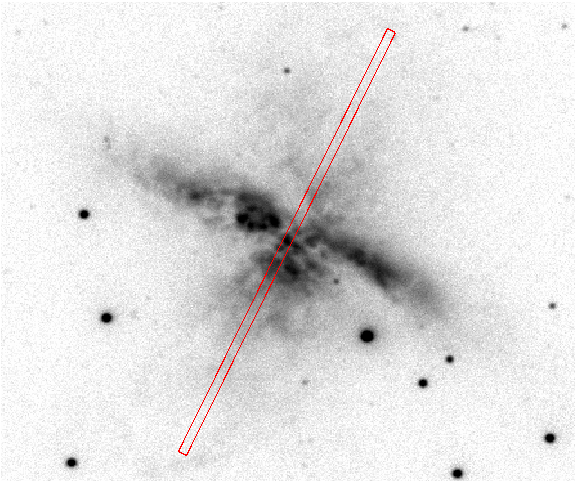}
}
\caption{GALEX FUV (left) and NUV (right) images of M82.  North is up and East is to the left.  The red box on each image is our extraction aperture.  Discussion of other extraction apertures and the implications for our results is provided in Section \ref{sec:discussion}.  M82 is at an inclination angle of $80^{\circ}$ such that the southeast portion of the wind is tilted toward us.}
\label{fig:M82im}
\end{figure*}

We examine the particular case of M82 (distance 3.63~Mpc, Gerke et al.\ 2011, Freedman et al.\ 1994), one of the nearest starburst galaxies to the Milky Way, in order to test whether radiation pressure is important in this case.  There are extensive archival observations of M82 in many bands (see, e.g., McLeod et al.\ 1993, Westmoquette et al.\ 2007).  However, the starburst core of M82, where most of the UV and optical radiation that would be important for driving the wind is generated, is heavily obscured by dust along our line of sight because M82 is nearly edge-on (inclination:  $\mbox{i}=80^{\circ}$, McKeith et al.\ 1993).  The wind itself is quite bright in UV light on kpc scales above the plane (Figure~\ref{fig:M82im}), and work by Hoopes et al.\ (2005) indicates that this emission is mostly due to dust in the wind scattering UV light from the central starburst into our line of sight, rather than other potential sources of \textit{in situ} UV emission such as photoionization and shocked gas.

The fundamental measure of whether radiation pressure is currently important in driving the wind is the Eddington ratio, $\Gamma = L_{\star}/L_{\rm Edd}$.  To calculate $\Gamma$, we need to know the dynamical mass of M82 interior to our observation point a height $z$ above the plane, $M(<z)$, the flux-mean opacity of the dust/gas wind fluid $\kappa$, and the true luminosity of the starburst that escapes to a height $z$ from the disk, $L_{\star}$.  

Although the bolometric luminosity of M82 is fairly well known from FIR data ($L_{\rm IR} = 5.9\times 10^{10}\:{\rm L_{\odot}}$, Sanders et al.\ 2003), the shape of the SED in the UV above the plane, and specifically the UV luminosity $L_{\rm \star,\,UV}$ at a height $z$ above the plane, is unknown without modeling.  In principle, $L_{\rm \star,\,UV}$ could be as high as $L_{\rm IR}$.  If so, assuming spherical symmetry,
\begin{equation}
L_{\rm Edd}= \frac{4\pi GcM_{\star}}{\kappa} = 1.3\times 10^{11}\:{\rm L_{\odot}}\,M_{10}\,\kappa_3 ^{-1}
\end{equation}
where $M_{10}=M_{\star}/10^{10}\:{\rm M_{\odot}}$ and $\kappa_3 = \kappa/10^3\:{\rm cm^2\: g^{-1}}$ of gas, normalized to a value appropriate for an SED dominated by UV light (Laor \& Draine\ 1993).  The Eddington ratio would then be $\Gamma = L/L_{\rm Edd} \sim 0.5$, suggesting that the radiation pressure force on the dusty gas could be comparable to gravity if the UV luminosity above the plane is of order $L_{\rm IR}$.  For comparison, the total observed UV luminosity of M82 in the GALEX bands from Hoopes et al.\ (2005) is $1.2\times 10^8\:{\rm L_{\odot}}$, or $\approx 2\times 10^{-3}L_{\rm IR}$ and $\approx 9\times 10^{-4}L_{\rm Edd}$.

By modeling the dust scattering, we work backwards using archival GALEX images of M82, presented by Hoopes et al.\ (2005) (see Figure~\ref{fig:M82im}), to obtain the UV luminosity escaping perpendicular to the disk of M82.  We then model the stellar population, which together with the scattered UV flux, allows us to obtain an SED of the radiation field at height $z$ and to calculate the Eddington ratio with the chosen dust model.  This calculation includes an MRN distribution (Mathis et al.\ 1977), STARBURST99 spectra (Leitherer et al.\ 1999), and dust grain properties calculated by Laor \& Draine\ (1993).  Changes in the grain size distribution are discussed in Section~\ref{sec:results}.

In Section~\ref{sec:estimate} we give an order of magnitude estimate of the UV luminosity that escapes perpendicular to the plane, as well as estimates of various parameters.  In Section~\ref{sec:results} we provide our results.  In Section~\ref{sec:error} we give a description of the uncertainties in our work. A brief discussion and summary are presented in Sections~\ref{sec:discussion} and \ref{sec:summary}.

While we were completing this work, Socrates \& Sironi\ (2013) also presented a calculation of the Eddington ratio for M82 (top panel of their Fig.~1).  However, their calculation makes use of the SED model of Silva et al.~(1998), which makes no correction for the inclination dependence of the SED and the different shape of the radiation field emerging perpendicular to the disk.  Our work is distinct in that we account for the inclination dependence, estimate the SED shape from the scattered FUV and NUV light, and estimate $\kappa$ along the minor axis.  In addition, by taking into account the observed rotation curve of the M82, we are able to estimate the Eddington ratio as a function of height along the minor axis.  We are then able to make an estimate of the single-scattering Eddington limit within the starburst (Section~\ref{sec:discussion}).

\section{Estimates} \label{sec:estimate}

\subsection{Order of Magnitude Calculation}

We expect that $L_{\rm \star,\,UV} < L_{\rm bol}$ because at least a portion of the UV radiation in the starburst is absorbed within the system itself and never emerges to kpc scales above the plane.

In order to estimate the UV flux which escapes the starburst region and scatters on grains a distance $z$ above the plane, we make several simplifying assumptions:  first, that the starburst disk is a point source; second, that the dust in the wind outside of the starburst region is optically thin to the scattered UV (i.e., each photon scatters at most once); third, that the density of the wind is constant along our line of sight at a given height above the plane of M82; and fourth, that the wind is illuminated as a right circular cone.  Treating this system as steady-state, we can write down the radiative transfer equation for this system:
\begin{equation} \label{eq:rad}
\frac{dI_{\lambda}}{ds} \approx \frac{\alpha_{\lambda}}{4\pi} \frac{L_{\star,\,\lambda}}{4\pi r^2} \rho \kappa_{\rm s,\,\lambda} \: ,
\end{equation}
where $L_{\star,\,\lambda}$ is the specific starburst luminosity at wavelength $\lambda$, $r$ is the distance from the central source, $\rho$ is the density of the dust, $\kappa_{\rm s}$ is its scattering opacity, and ${\alpha}$ is a wavelength-dependent factor that captures the anisotropy of dust scattering in the UV.

\begin{figure}[t]
\centering
\includegraphics[scale=0.6]{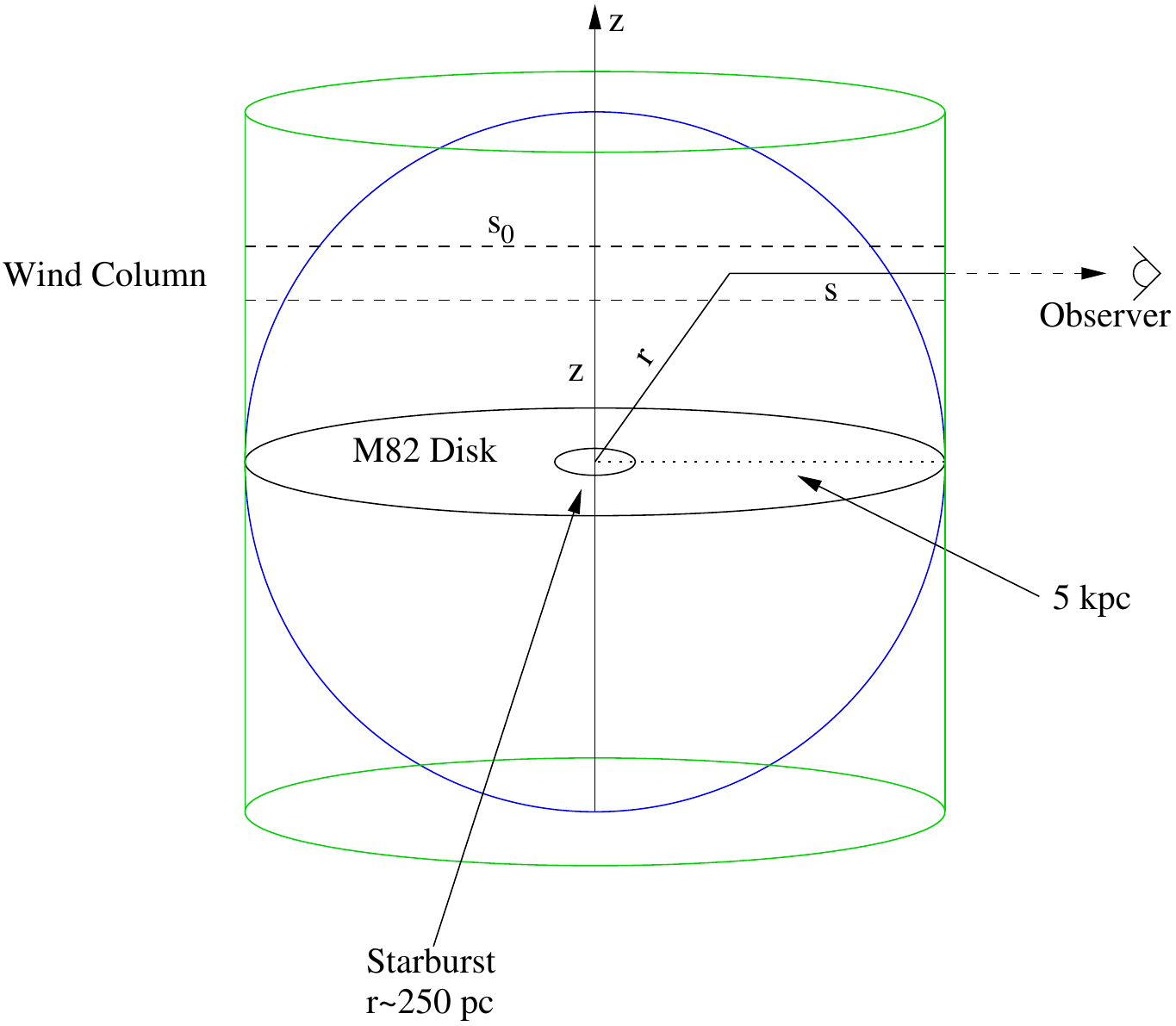}
\caption{Diagram of each of our model geometries for the dust distribution.  Blue shows the spherical model, and green the cylindrical model.  $r$ is the distance from the center of M82 to a given portion of the wind, $s$ is the path length through the dust model to the observer from that portion, $s_0$ the total path length through the illuminated portion of the dust model, and $z$ the distance from the midplane of M82, as defined in Section~\ref{sec:estimate}.  The starburst core and extended optical disk of M82 (${\rm i} = 80^{\circ}$) are also shown.  Not to scale.}
\label{fig:models}
\end{figure}

Because the total path length through the wind is comparable to the height above the disk, $r$ changes significantly when integrating over a path through the wind that runs parallel to the disk.  Along such a path (see Figure~\ref{fig:models}), 
\begin{equation}
r = \sqrt{\left(s - \frac{s_o}{2} \right)^2 + z^2}\: ,
\end{equation}
where $s_o$ is the total thickness of the illuminated portion of the wind, $s$ is the path length actually traversed by a given photon on this path, and $z$ the height above the disk.  This gives
\begin{equation}
I_{\lambda} = \int _0 ^{s_o} \frac{L_{\star,\,\lambda}\alpha_{\lambda}\rho\kappa_{\rm s,\,\lambda} ds}{16\pi^2 \left[z^2 + \left({s_o}/{2} - s \right)^2 \right] } = \frac{L_{\star,\,\lambda}\alpha_{\lambda}\rho\kappa_{\rm s,\,\lambda}}{8\pi^2 z} \tan^{-1} \left({\frac{s_o}{2z}}\right)\: .
\end{equation}
The observed flux from a given aperture in the M82 wind is then
\begin{equation}
F_{\rm obs,\,\lambda} = \int I_{\lambda} \cos\theta d\Omega = 2\pi \int _0 ^{\theta_{\rm c}} I_{\lambda} \cos\theta \sin\theta d\theta \: ,
\end{equation}
where the angle $\theta_{\rm c}$ is the angular radius of the source aperture.  For our work, we use an aperture with a radius of five GALEX pixels, or $7\farcs 5$ (compare to GALEX's resolution of $4\farcs 2$ and $5\farcs 3$ in the FUV and NUV, respectively).  Substituting, one finds an expression for the observed flux, which can then be inverted to give the luminosity of the starburst at wavelength $\lambda$ escaping the central starburst region
\begin{equation} \label{eq:Lstar}
L_{\star,\,\lambda} = \frac{8\pi z F_{\rm obs,\,\lambda}}{\alpha_{\lambda}\rho\kappa_{\rm s,\,\lambda}\tan^{-1}(s_o/2z)\sin^2\theta_{\rm c}} \: .
\end{equation}

The GALEX images indicate that the dust is illuminated as a cone with a $\sim 55^{\circ}$ opening angle, so that $s_o \approx z$. With reasonable values of the other parameters, the luminosity is
\begin{equation} \label{eq:Lstarnum}
\left. \lambda L_{\star,\,\lambda} \right|_{\rm FUV}= 1.4\times 10^{42}\frac{z_{\rm kpc}\lambda_{1516}F_{\rm obs,-17}}{\alpha_{.35}\rho_{-28}\kappa_{\rm s,4.6}} \:{\rm ergs\:s^{-1}} \: ,
\end{equation}
where $z_{\rm kpc}=z/{\rm kpc}$ is the distance above the plane of M82, $\alpha_{.35}=\alpha /0.35$ is a geometric factor that captures the physics of anisotropic UV scattering off dust grains, $\rho_{-28} = \rho/10^{-28}\:{\rm g\: cm^{-3}}$ is the mass density of the dust distribution, $\kappa_{\rm s,4.6} = \kappa_{\lambda}/10^{4.6}\:{\rm cm^{2}\: g^{-1}}$ is the scattering opacity per gram of dust, $F_{\rm obs,-17} = F_{\rm obs}/10^{-17}\:{\rm ergs\:s^{-1}\:cm^{-2}}\:\text{\AA}^{-1}$ is the observed flux, and $\lambda_{1516} = \lambda/1516\:\text{\AA}$ is the effective wavelength of the FUV band.

The sum of the UV luminosity in the two GALEX bands is $L_{\rm \star,\,UV}= \Delta\lambda_{\rm FUV}L_{\rm \star,\,FUV} + \Delta\lambda_{\rm NUV}L_{\rm \star,\,NUV} \approx 2\times 10^{42}\:{\rm ergs\:s^{-1}}$, where $\Delta\lambda$ is the effective bandwidth of the GALEX filters (268~\AA ~for FUV and 732~\AA ~for NUV).  This is $\sim$5 times the total observed UV luminosity from Hoopes, et al.\ (2005) and about $10^{-2}L_{\rm IR}$.

\subsection{Parameter Estimates} \label{sec:param}

In order to estimate the density distribution of dust in the superwind, we use the dust column mass map from Figure~2 of Roussel et al.\ (2010).  This dust map is based on Herschel SPIRE $250\:\mu{\rm m}$ and $500\:\mu{\rm m}$ data. Roussel et al.\ (2010) used these data to calculate a map of the dust mass surface density, $\Sigma$, with a resolution of 170 pc. They estimate that the total dust mass in the halo is $\sim 10^{6}\: {\rm M_{\odot}}$ and at least 65\% of the dust is not associated with the superwind.  From this map, we get a dust column density for each location in the wind.

\begin{figure}[t]
\centering
\includegraphics[scale=0.45]{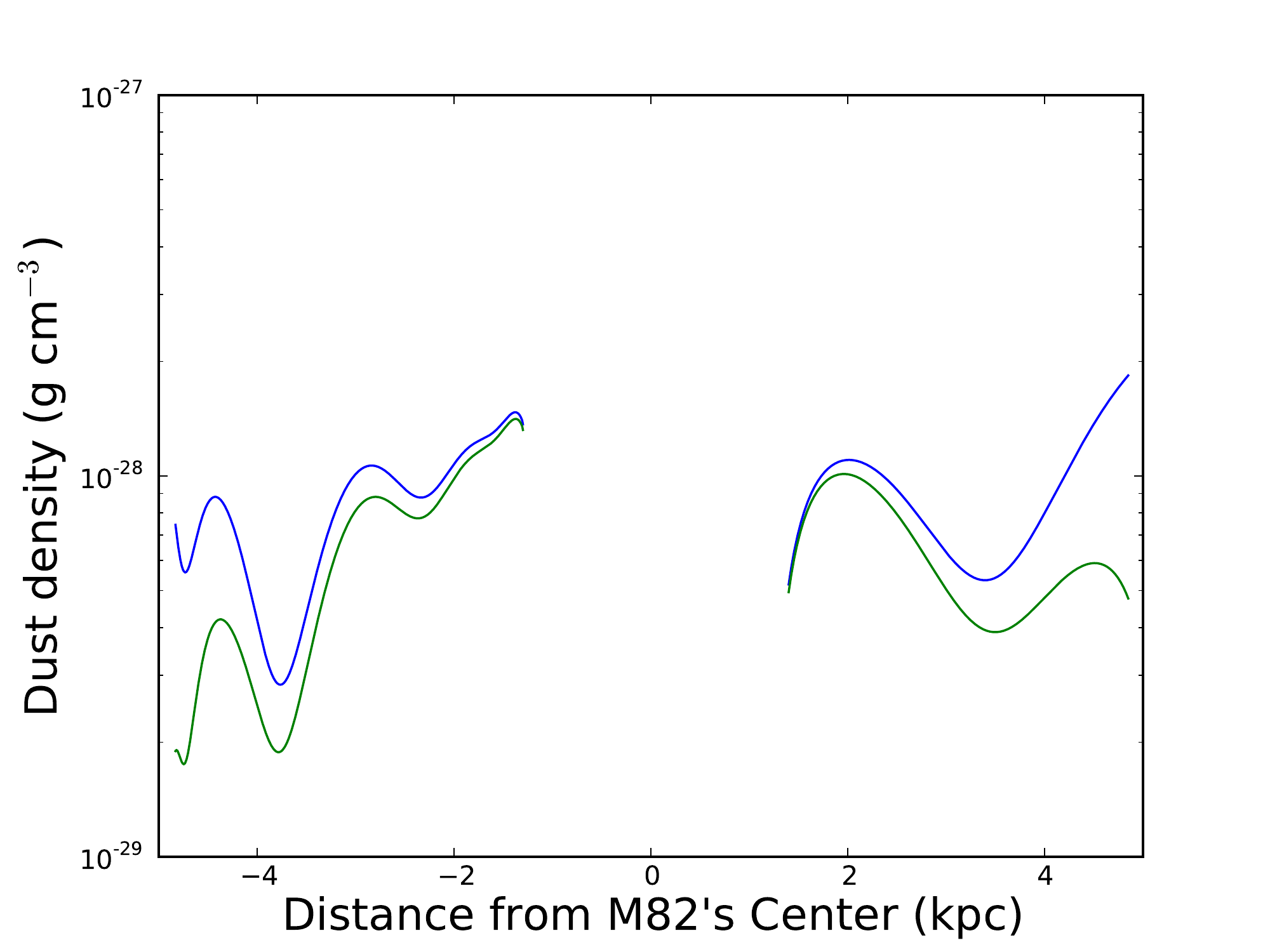}
\caption{Dust density calculated from Roussel et al.\ (2010) for each of our model geometries along the extraction aperture shown in Figure~\ref{fig:M82im} along the minor axis of M82 (positive and negative are north and south, respectively).  The green curve is for the cylindrical model, and blue for the spherical (see Figure~\ref{fig:models}).  The central region is empty because the Roussel et al.\ (2010) map provides no data there (see their Figure~2).  We applied a smoothing spline to the data for the figure.  We also tested different extraction apertures, but found that using a different region of the wind had no significant effect on our results (Section~\ref{sec:error}).}
\label{fig:dust}
\end{figure}

We assume two different model geometries: a spherical distribution truncated at 5~kpc and a 5~kpc radius cylinder, both centered on M82 (Figures~\ref{fig:models} and \ref{fig:dust}).  We then assume that the volume density of dust in each column is constant, so that $\rho = \Sigma / s_o$. In Figure~\ref{fig:dust} we show that both of these geometries produce dust density estimates within an order of magnitude of $10^{-28}\:{\rm g\: cm^{-3}}$ (Equation~\ref{eq:Lstarnum}).  The overall shapes of the different distributions are the result of the behavior of $s_o$; the spherical model produces higher densities further out because $s_o$ rapidly shrinks at the extremities.  It should be noted that the UV light from the starburst illuminates only a small portion of either of these models (Figure!\ref{fig:M82im}).  This is consistent with the observed morphology of the UV and FIR maps, since the latter shows an extended dust distribution, while the UV images show a smaller conical illuminated region.

To calculate the dust opacity, we assume an MRN dust distribution (Mathis et al.\ 1977), with 50\% each of graphite and silicate grains.  We then use the absorption and scattering cross sections from Laor \& Draine\ (1993).  We combine this with unreddened STARBURST99 models of the stellar spectra (Leitherer et al.\ 1999), and calculate a total flux-mean scattering opacity integrated over the entire SED of $\kappa_{\rm s,F} \sim 4\times 10^4\:{\rm cm^{2}\: g^{-1}}$ of dust (Equations~\ref{eq:rad}, \ref{eq:Lstarnum}), assuming a bimodal stellar population with components aged 4 and 9~Myr (F\"{o}rster Schreiber et al.\ 2003).

\begin{figure}
\centering
\includegraphics[scale=0.45]{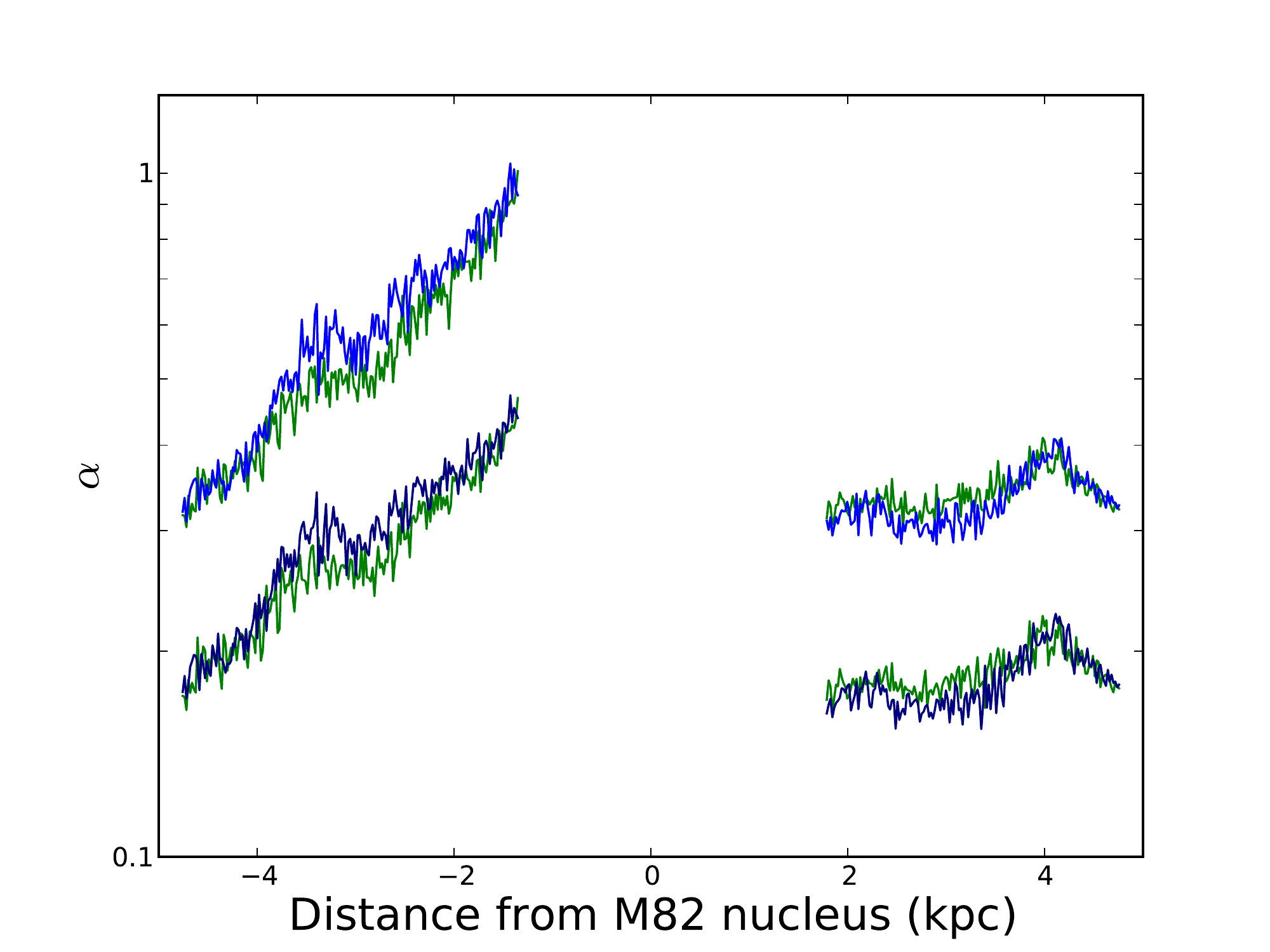}
\caption{$\alpha$ as a function of $z$.  The darker curves are for NUV, the brighter for FUV.  The blue curves correspond to the spherical geometry, and the green the cylindrical.}
\label{fig:alpha}
\end{figure}

\begin{figure*}[t]
\centerline{
\includegraphics[scale=0.45]{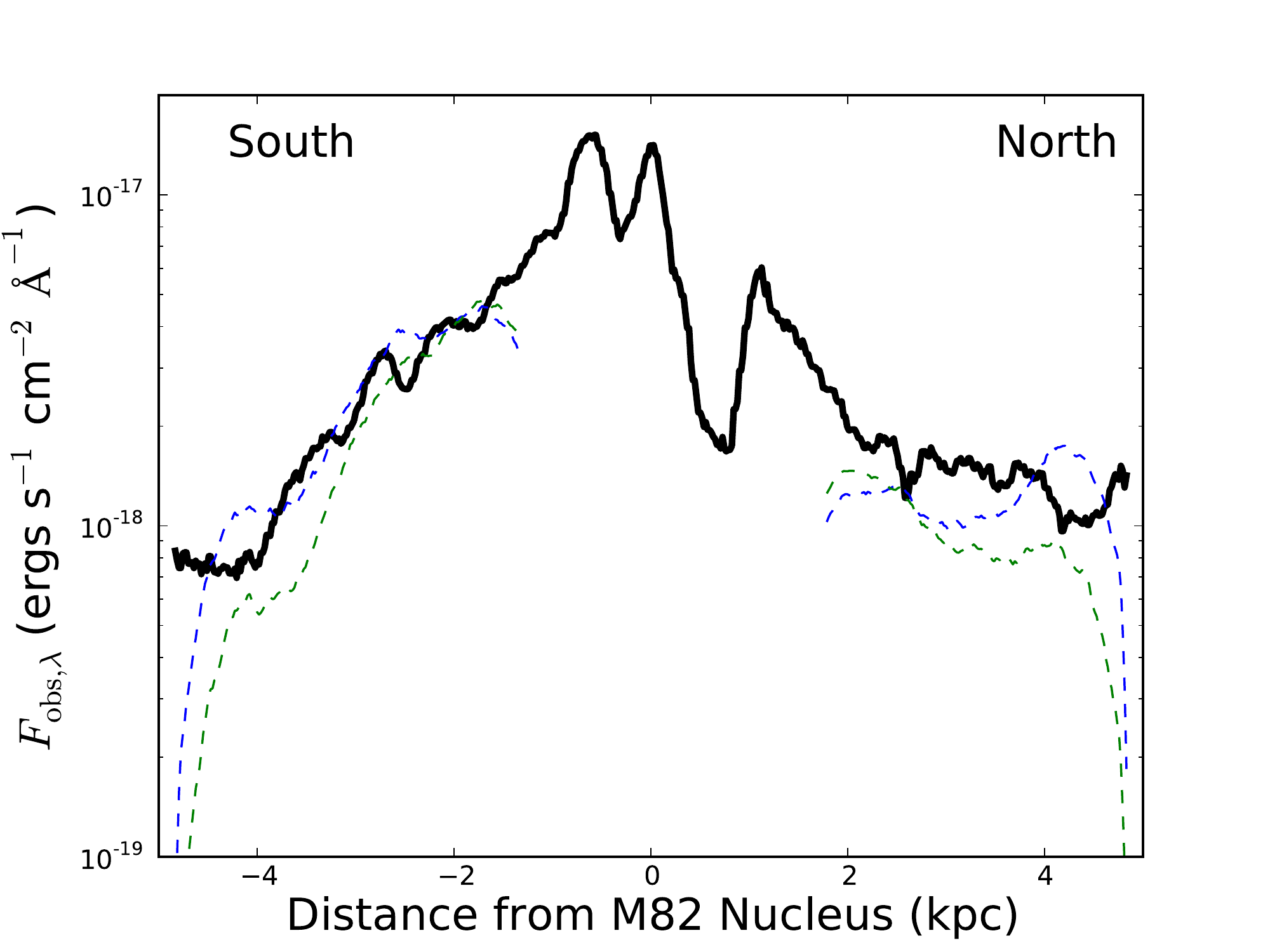}
\includegraphics[scale=0.45]{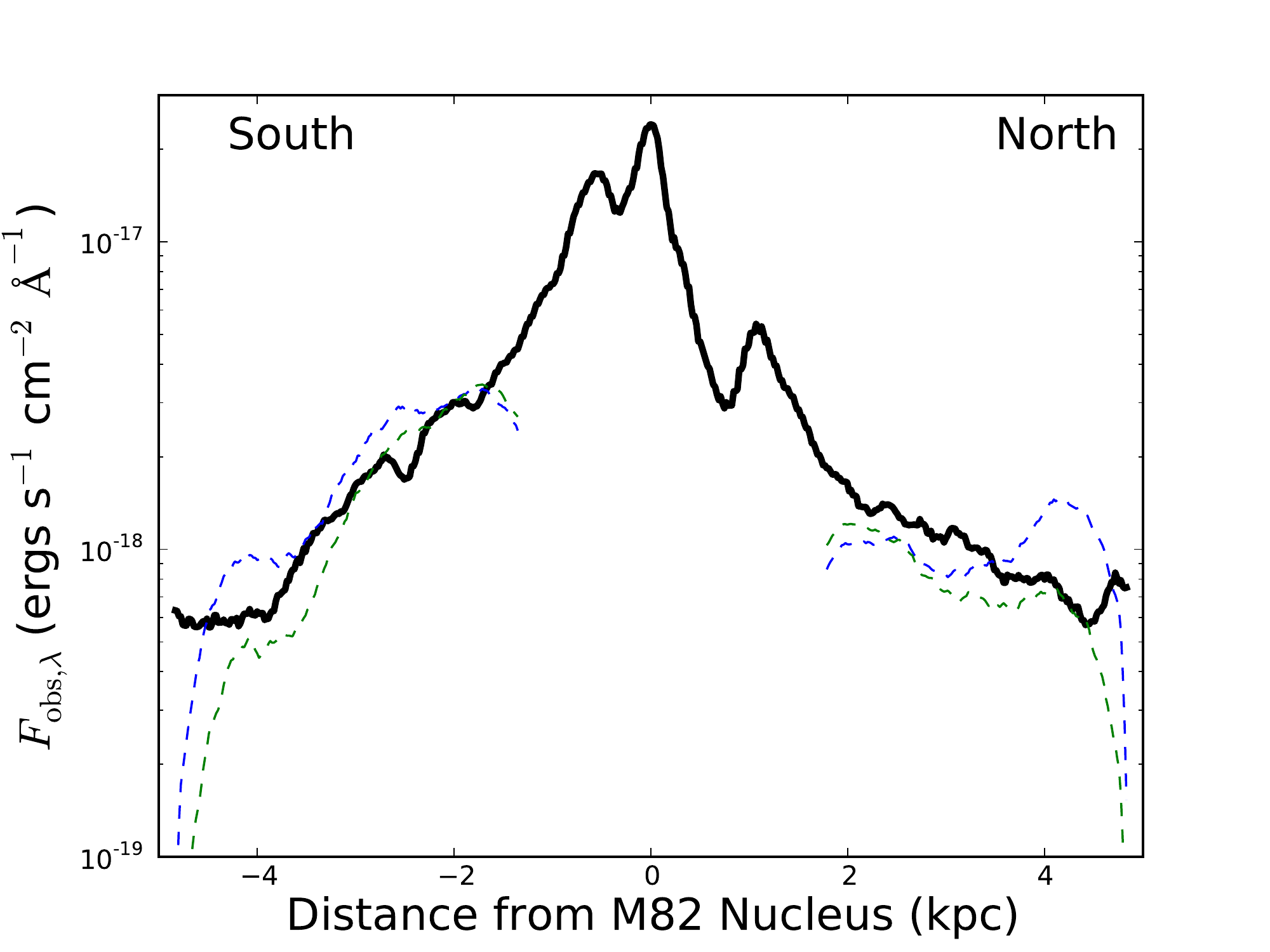}
}
\caption{Observed flux as a function of projected distance from M82 for the GALEX FUV (left) and NUV (right) data and our simulations.  We used a circular aperture five GALEX pixels, or $7\farcs 5$, in radius to calculate the average observed flux, $F_{\rm obs, \lambda}$, at each point.  Negative values on the x-axis are for areas south of the plane of M82, while positive ones are to the north.  The black curve is the GALEX data (Figure~\ref{fig:M82im}).  The dashed green and blue curves show results from our Monte Carlo simulations assuming the dust density distribution from Roussel et. al\ (2010), calculated assuming the cylindrical and spherical dust geometries, respectively, shown in Figure~\ref{fig:models} and \ref{fig:dust}.  Each of the simulation curves are normalized to the GALEX data at 2~kpc south of the plane.  The southern portions of the curve are brighter due to the anisotropy in the scattering.  The hole in the middle corresponds to the hole in the Roussel et al.\ (2010) dust density data; we simply set the density in that region to zero.  The rapid falloff at the outer edges occurs for much the same reason; the simulation region abruptly cuts off at 4.8~kpc.}
\label{fig:sim}
\end{figure*}

We use three dimensional Monte Carlo scattering simulations to estimate $\alpha$. Each simulation run (one for each model geometry and bandpass) used $5\times10^7$ photons with a weighting scheme that guaranteed each photon scattered once (Wood et al.\ 2001).  We assumed that the wind was optically thin, and that attenuation due to absorption was minimal and could be ignored.  To substantiate that assumption, we calculated the optical depth along each sight line.  In the region for which we have dust density data, the total optical depth in either band for an average sight line is $\sim 0.05$ in the FUV, given the Roussel et al.\ (2010) dust density data and our model geometries.  We ignored light from M82's outer disk, and approximated the starburst as a point-source.  Modeling the starburst as a disk of radius 250~pc (F\"{o}rster Schreiber et al.\ 2003) is a $\sim5\%$ effect at 1~kpc from the plane, and becomes less important at larger distances. As shown in Figure~\ref{fig:dust}, we have no dust density data within $\sim 1300$~pc of the center of M82, because the Roussel et al.\ (2010) map provides no estimate of $\Sigma$ in that region due to difficulties in subtracting the disk PSF.  For the purposes of the simulation, we set the density to zero there, allowing photons to propagate freely outward until they encounter the overlying wind material.  This hole in the dust distribution results in zero scattering events into our line of sight.  We ultimately removed the region from 1.5~kpc south to 1.8~kpc north (see Figures~\ref{fig:sim} and \ref{fig:Edd}).

The geometric factor $\alpha$ is the ratio of the predicted flux with the Draine\ (2003a) phase function to the predicted flux for the case of isotropic scattering.  Figure~\ref{fig:alpha} plots $\alpha$ as a function of distance from the plane of M82.  For the FUV band, $\alpha \sim 0.35$ and for the NUV band $\alpha \sim 0.18$ for both of our model geometries for the northern lobe of the wind.  Due to the fact that the southern lobe is pointed towards us, $\alpha$ varies by a factor of $\sim$5 over that portion of the wind, from $\sim$0.3 at 4.5~kpc to $\sim$1.5 at 1.5~kpc for the FUV band.  This difference in the behavior of $\alpha$ between the northern and southern lobes arises from the shape of the Draine phase function.  The Draine phase function is relatively flat at the scattering angles covered by the northern lobe of the wind, but begins to rise more steeply as the dust moves into the forward-throwing regime.

\section{Results} 
\label{sec:results}

\subsection{The Intrinsic UV Luminosity}
\label{section:fobs}

We measured the observed flux from the wind along a linear aperture perpendicular to the plane of M82 that extends 5~kpc north and south of $9^{\rm h}55^{\rm m}52^{\rm s}.7,\:69^{\circ}40'46''$ (Jackson et al.\ 2007). At each point along the line, we then measured the flux within a circular aperture five GALEX pixels in radius (GALEX's resolution is $\sim 3$ pixels).  The average flux in each aperture and in each band was calculated and then corrected for  Galactic extinction ($E(B-V)=0.159$; Schlegel et al.\ 1998) in the GALEX bands (Wyder et al.\ 2007).  The resulting flux in both bands is $F_{\rm obs}$ (see Equation~\ref{eq:Lstar}) and is shown as the solid lines in the two panels of Figure \ref{fig:sim}.

The simulation results for the two adopted geometries are plotted against our extraction aperture through the GALEX data in Figure~\ref{fig:sim}. Aside from an overall normalization applied such that all curves intersect at 2~kpc south, we have not tried to fit the GALEX observations.  These curves use the two assumed wind geometries (Figure~\ref{fig:models}) together with the Roussel et al.\ (2010) dust data (Figure~\ref{fig:dust}).  The overall difference in the flux between the northern and southern parts of the simulation results is due to the anisotropy in the scattering: the southern lobe of the wind is pointed towards us (${\rm i} = 80^{\circ}$), and the dust is forward-throwing in the UV.  

\begin{figure}[t]
\centering
\includegraphics[scale=0.45]{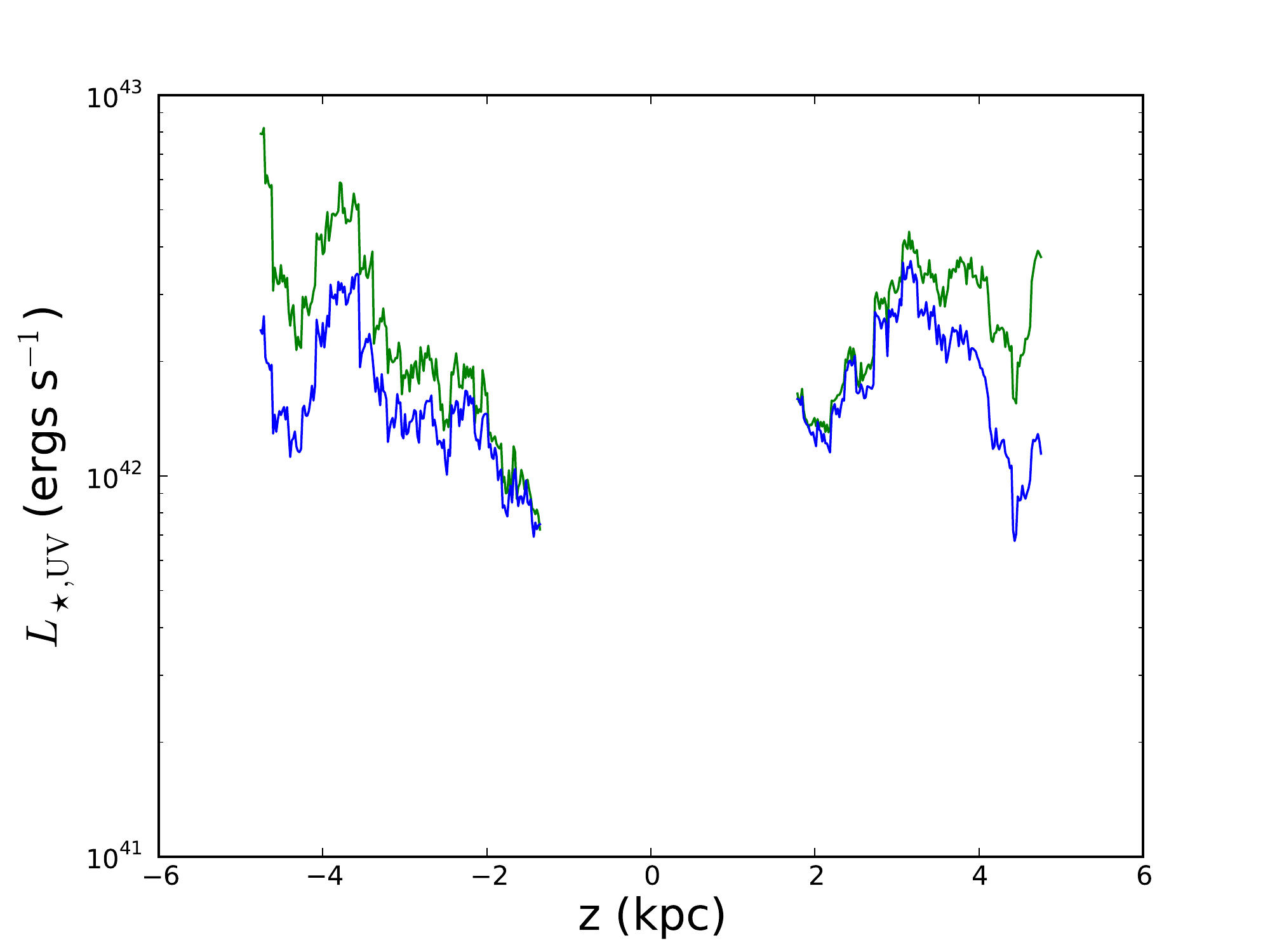}
\caption{The total UV luminosity in the GALEX bands escaping the starburst region ($L_{\star,\,\rm UV}$) as determined from Equation~\ref{eq:Lstarnum} in both model geometries (same as Figure~\ref{fig:models}), incorporating $\alpha(z)$ from the Monte Carlo scattering calculations, $\rho(z)$ from Figure~\ref{fig:dust}, and the UV surface brightness from Figures~\ref{fig:M82im} and \ref{fig:sim}.  If the geometry was correct and the dust density distribution perfectly determined, we would expect the inferred value of $L_{\star,\,\rm UV}$ to be constant as a function of $z$.}
\label{fig:Lstar}
\end{figure}

Combining our model geometries, scattering calculations (for $\alpha(z)$; see Section \ref{sec:param}), and the Roussel et al.\ (2010) dust data then gives $L_{\star, \rm FUV}$ and $L_{\star, \rm NUV}$ (Equation~\ref{eq:Lstar}), which we combine to give $L_{\star, \rm UV}$, the escaping luminosity in the GALEX bands, at each $z$.  We plot our inferred $L_{\star, \rm UV} (z) $ for each model geometry in Figure~\ref{fig:Lstar}.  Given the assumptions of our model, $L_{\star, \rm UV} (z)$ should be constant as a function of height since only one total UV flux escapes the starburst region and there is minimal absorption perpendicular to the disk outside of the starburst region.  The fact that $L_{\star, \rm UV}$ as calculated is not constant with $z$ indicates that our model fails to capture the true geometry of the dust distribution.  In principle, one could use the constraint that  $L_{\star, \rm UV}(z)=$\,constant to determine (or constrain) the true dust density distribution along the line of sight at every $z$.  A future work could iterate between the scattering calculations and the dust distribution using the joint constraints of $F_{\rm obs}$, the Roussel et al.\ data, and $L_{\star, \rm UV}(z)=$\,constant to determine the true dust distribution.  

For our purposes, it is sufficient to take the range of $L_{\star, \rm UV}(z)$  from the spherical and cylindrical geometries as a measure of our systematic errors and we find that $L_{\star, \rm UV}(z)\simeq1-6\times10^{42}$\, ergs s$^{-1}$, a factor of $\simeq2-12$ times larger than the total UV luminosity inferred by Hoopes et al.~(2005) from our line of sight.

\subsection{The UV$-$NIR SED}
\label{section:sed}

\begin{figure}[t]
\centering
\includegraphics[scale=0.45]{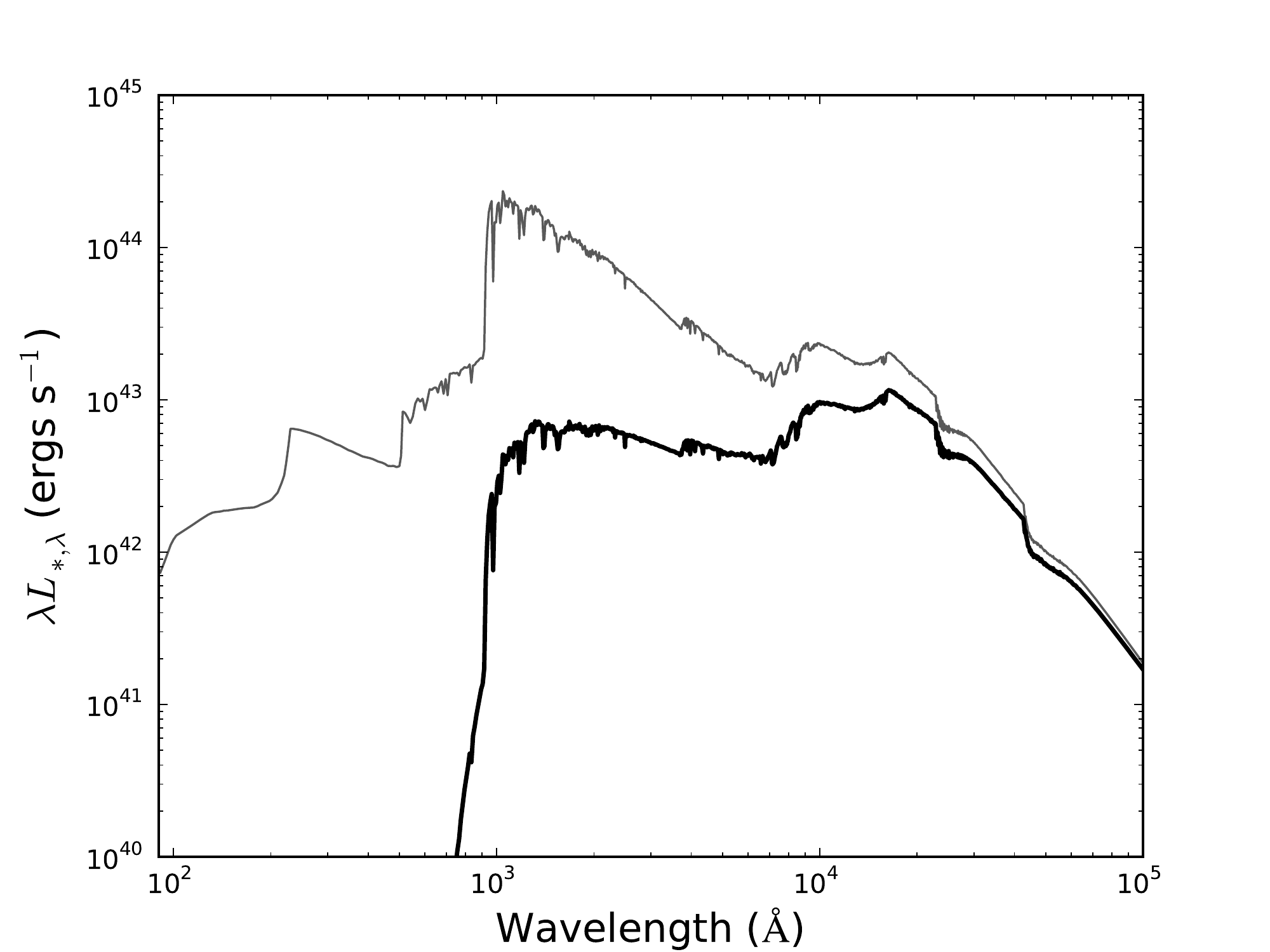}
\caption{The solid gray line shows the underlying assumed SED of the stellar population, as determined from the population synthesis modeling of F\"{o}rster Schreiber et al.\ (2003).  The black solid line shows an example of the extincted model used to calculate the Eddington ratio in Equation~\ref{eq:Edd} and shown in Figure \ref{fig:Edd}. This model gives $L_{\rm \star,\,UV}=2\times10^{42}$\,ergs s$^{-1}$, which corresponds to $\tau_{\rm FUV}\approx3$ for a foreground screen.  Given each of the derived values of $L_{\star,\,\rm UV}(z)$ in Figure \ref{fig:Lstar} we choose $\tau_{\rm FUV}$ appropriately, and then use the SED model to calculate
$\Gamma(z)$.}
\label{fig:SED}
\end{figure}

We next constrain the flux in the optical and NIR escaping perpendicular to the disk by comparing $L_{\star, \rm UV}$ with the true unreddened GALEX band luminosities expected from the work of F\"{o}rster Schreiber et al.\ (2003), who modeled the stellar population in the central starburst region from NIR spectroscopy (F\"{o}rster Schreiber et al.\ 2001). We produced STARBURST99 models for instantaneous bursts of star formation corresponding to a combination of a 4 and 9 Myr-old starbursts with parameters given by F\"{o}rster Schreiber et al.\ (2003): $L_{\rm bol}=6.6\times 10^{10}\:{\rm L_{\odot}}$ and $M_{\star} = 6.1 \times 10^8\:{\rm M_{\odot}}$ for the combined bursts at the current epoch. We weight the STARBURST99 spectrum by the GALEX response curves for each band and calculate the GALEX FUV and NUV band luminosities of this fiducial, unextincted STARBURST99 model.  As shown in Figure~\ref{fig:SED}, the unreddened UV luminosity is $\sim2\times10^{44}$\,ergs s$^{-1}$.  Our derived UV luminosities are much smaller ($1-6\times10^{42}$\,ergs s$^{-1}$; Figure \ref{fig:Lstar}), indicating that a significant fraction of the UV luminosity of the starburst is absorbed on small scales.

We model the absorption of the starburst UV flux using a simple foreground screen of dust.  The natural logarithm of the ratio of our derived $L_{\star, \rm FUV}$ to the fiducial STARBURST99 GALEX band luminosity then yields $\tau_{\rm FUV}$.  Since $L_{\star, \rm FUV}$ (or NUV, or combined UV) varies as a function of $z$ (Fig.~\ref{fig:SED}; Section \ref{section:fobs}), we derive a different optical depth at every $z$.  We find that $\tau_{\rm FUV}$ ranges from 1.6 to 4, with the larger values corresponding to the lower $L_{\star, \rm UV}$ in Figure \ref{fig:SED} and vice versa.

To determine the optical and NIR luminosities emerging from the starburst region, we use the extinction law from Calzetti et al.\ (1994) and adjust it to provide our calculated $\tau_{\rm FUV}$ at 1516~\AA, the effective wavelength of the FUV GALEX band.  We then apply the Calzetti et al.\ law to our fiducial STARBURST99 SED, producing a reddened spectrum which we use for our Eddington ratio calculation. In Figure~\ref{fig:SED}, we show an example of how this reddened spectrum compares to the fiducial SED for one choice of $\tau_{\rm FUV}$.

\begin{figure*}[t]
\centering
\includegraphics[scale=0.7]{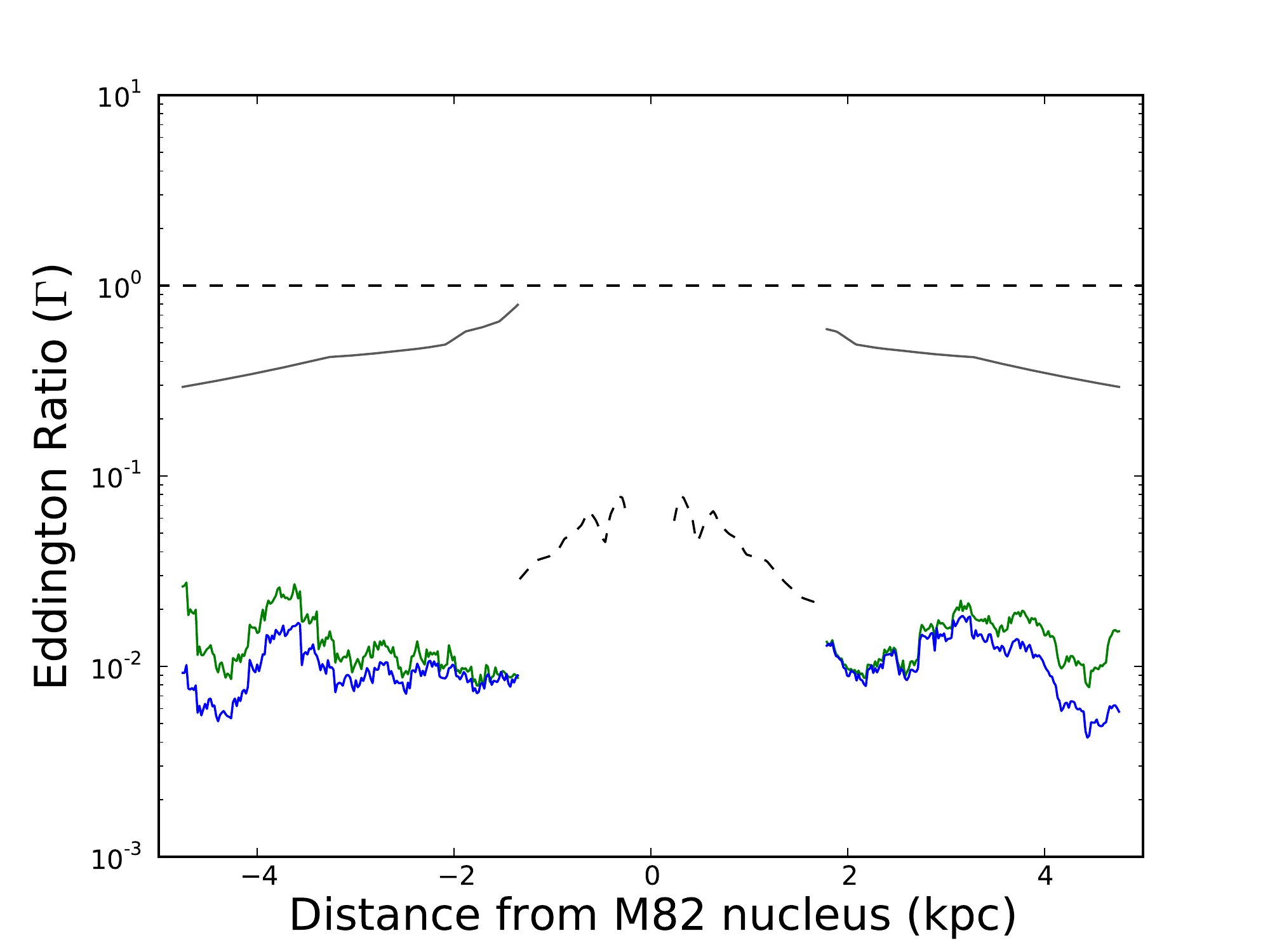}
\caption{Eddington ratio as a function of distance north of the plane of M82 as calculated from Equation~\ref{eq:Edd}.  The green curve uses the cylindrical geometry, and the blue curve the spherical.  All curves use the FUV data only; the curves generated from the NUV data are not significantly different.  The gray curve shows the Eddington ratio calculated using the fiducial STARBURST99 SED with no extinction.  The dashed black curve extends our calculation inward using a constant $L_{\star, \rm UV} \approx 2\times 10^{42}\:{\rm ergs \: s^{-1}}$ for $\tau_{\rm FUV} = 3$.  The black dashed and grey curves are relatively smooth because they keep $L_{\star, \rm \lambda}$ constant with $z$.  If we employ the full range of values implied for $L_{\star, \rm UV}$ in Figure~\ref{fig:Lstar}, the maximum deviation in the blue and green curves is a factor of six, from 0.01 - 0.06, and the maximum range in the dashed curve goes as high as $\sim 0.2$.  The middle region is not shown for the colored curves due to the hole in the Roussel et al.~(2010) dust map (see Figures~\ref{fig:dust} and \ref{fig:sim}), as explained in Section~\ref{sec:param}.  See Section~\ref{sec:discussion} for more discussion of this region.}
\label{fig:Edd}
\end{figure*}

\subsection{The Eddington Ratio}

The Eddington ratio for the combined dust and gas wind fluid, integrated over an MRN distribution, the redenned STARBURST99 spectral model, and using a thin, uniformly bright disk for the geometry of the starburst and a spherically symmetric mass distribution for the disk and galaxy, is
\begin{equation} \label{eq:Edd}
\begin{split}
\Gamma = {} & \frac{Ar^2}{4cGM_{\star}m_{\rm H}(R_{\rm s}^2+r^2)} \\
& \times \int a^{\beta+2}L_{\star,\,\lambda}[Q_{\rm a}+Q_{\rm s}(1-g)]\,da\,d\lambda
\end{split}
\end{equation}
where $A$ is the normalization of the MRN distribution (calculated via the method in Laor \& Draine\ 1993), $M_{\star}$ is the dynamical mass of M82 interior to $r$, $m_{\rm H}$ is the mass of a hydrogen atom, $r$ is the distance from the center of M82, $R_{\rm s}$ is the radius of the starburst region, $a$ is the radius of a dust grain, $\beta$ is the slope of the grain size distribution, $L_{\star,\,\lambda}$ is the specific luminosity from Equation~\ref{eq:Lstar}, $Q_{\rm a}$ is the absorption efficiency, $Q_{\rm s}$ the scattering efficiency, and $g = \langle\cos\theta\rangle$ the average of the cosine of the scattering angle.  It should again be noted that $L_{\star,\,\lambda}$ is calculated at every $z$ above the plane in Figure~\ref{fig:Edd} (Figure~\ref{fig:Lstar}), even though if the wind is optically thin, $L_{\star,\,\lambda}$ should be constant as a function of height above the plane (Section \ref{section:fobs}).  We also calculate $\tau_{\rm FUV}$ for each $z$ and use this in correcting for the extinction in M82's disk so that the SED varies as a function of distance from the disk (as illustrated in Figure\ \ref{fig:SED}).  Although the variation in  $L_{\star,\,\lambda}$ as a function of height is an artifact of our assumed dust geometries, it allows us to understand the range of systematic error in our calculation of $\Gamma(z)$. 

In order to determine $M_{\star}(r)$, we use the work of Greco et al.\ (2012) (GMT12), who measured the rotation curve of M82 on large scales.  We assume circular orbits and a spherically symmetric mass distribution to determine $M_{\star}(r)$.  These assumptions break down within $\sim0.5$\,kpc due to the influence of the bar, as detailed in GMT12 (see also Westmoquette et al.\ 2007, 2009, 2012).  

The solid colored lines in Figure~\ref{fig:Edd} show our results for the Eddington ratio as a function of distance from the plane of M82 assuming a dust-to-gas ratio of $f_{\rm dg}=0.01$. Factoring in our large uncertainties (Section~\ref{sec:error}), $\Gamma\sim0.01-0.06$ for our model geometries, indicating that radiation pressure does not presently drive the starburst wind on $z\gtrsim$\,kpc scales.  Note that if the dust and gas were not hydrodynamically coupled, the dust grains would be super-Eddington, with $\Gamma\rightarrow\Gamma/f_{\rm dg}\sim1-3$.  We return to this issue in Section \ref{sec:discussion}.

The Eddington ratio $\Gamma$ increases on smaller scales ($z\lesssim 1$\,kpc) because $M_{\star}(r)$ decreases while $L_{\star,\,\lambda}$ likely stays roughly constant until $z$ is comparable to the starburst's vertical scale height $\sim 30\:{\rm pc}$ (Wei\ss\ et al.\ 2005).  The dashed line in Figure \ref{fig:Edd} shows $\Gamma(z)$ on small scales assuming the GMT12 rotation curve and a single value for $L_{\star,\,\rm UV}=2\times10^{42}$\,ergs s$^{-1}$.  Section~\ref{sec:discussion} gives more discussion of $\Gamma$ on small scales.

\begin{figure}[t]
\centering
\includegraphics[scale=0.45]{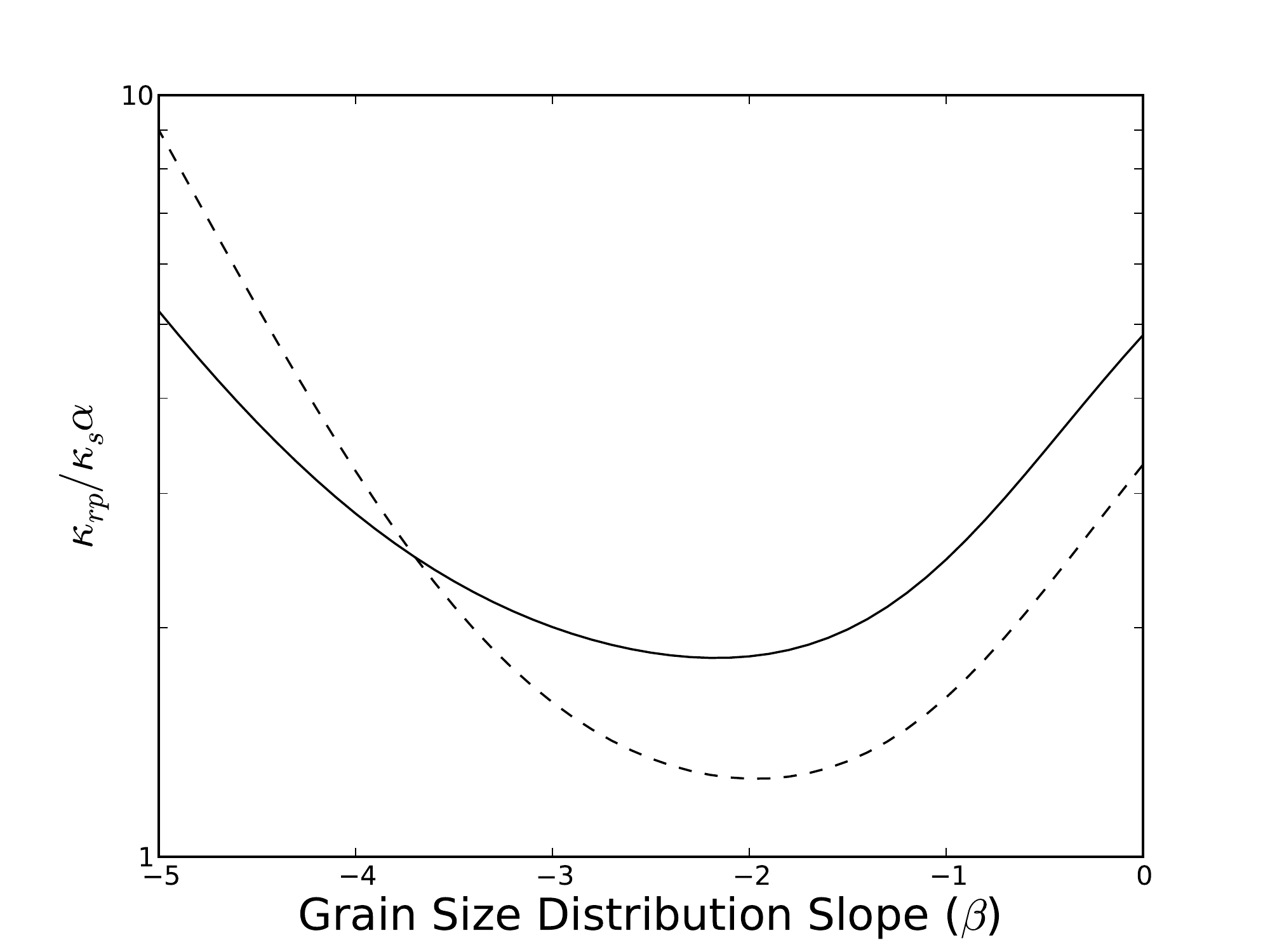}
\caption{Dependence of the Eddington ratio, $\Gamma$, on the slope of the grain size distribution, $\beta$.  The solid curve is for the FUV band, while the dashed curve is for the NUV band.  For an MRN distribution as used in this paper, $\beta=-3.5$.  $\beta=0$ is a completely flat grain size distribution.}
\label{fig:opacity}
\end{figure}

As seen in Equation~\ref{eq:Edd}, the functional form of $\Gamma$ depends heavily on the choice of grain size distribution and the optical properties of the grains.  The radiation pressure cross section, $Q_{\rm rp} = Q_{\rm a} + Q_{\rm s}(1-g)$, and the radiation pressure opacity, $\kappa_{\rm rp}$, are related by
\begin{equation}
n_{\rm d}\pi a^2 Q_{\rm rp} = \rho\kappa_{\rm rp}
\end{equation}
for spherical grains of radius $a$, where $n_{\rm d}$ is the number density of dust grains, $\rho$ is the dust mass density, and $\kappa_{\rm rp}$ is per gram of dust.  Since $Q_{\rm rp} \propto \kappa_{\rm rp}$, and $L_{\star,\,\lambda} \propto (\alpha_{\lambda}\kappa_{\rm s,\,\lambda})^{-1}$ (see Equation~\ref{eq:Lstar}), $\Gamma_{\lambda} \propto \kappa_{\rm rp,\,\lambda}/\alpha_{\lambda}\kappa_{\rm s,\,\lambda}$.  As the slope of the grain size distribution steepens, $\kappa_{\rm rp,\,UV}/\kappa_{\rm s,\,UV}$ grows; for an MRN distribution, $\beta=-3.5$ and $\kappa_{\rm rp,\,UV}/\kappa_{\rm s,\,UV}\approx 2.5$, while for $\beta=-4.5$, $\kappa_{\rm rp,\,UV}/\kappa_{\rm s,\,UV}\approx 4$, as shown in Figure~\ref{fig:opacity}.  For shallower grain size distributions, the opacity ratio changes by relatively small factors until the size distribution gets close to flat (Figure~\ref{fig:opacity}).  Holding the dust-to-gas ratio fixed, changing the size distribution does not change the dust mass calculated by Roussel et al.~(2010) by more than a few tens of percent.  This is because the FIR opacity is insensitive to the power law form of the grain size distribution at fixed dust-to-gas ratio, changing by only a few tens of percent from $\beta = -5$ to $\beta = -2$.  This means that over a reasonable range of slopes, the Eddington ratio does not change by more than a factor of three, too small to affect our overall result.

\section{Additional Uncertainties} \label{sec:error}

There are two main sources of uncertainty in our estimate of the gravitational acceleration on the dust grains:  the actual mass pulling on them, and the fact that the mass distribution is not perfectly spherically symmetric.  The first is a relatively small uncertainty, as GMT12 determined the total dynamical mass of M82 within 4 kpc from the center to within about 10\%.  Treating M82 as a thin uniform disk with $M=10^{10}\:{\rm M_{\odot}}$ and radius 1~kpc instead of a sphere/point mass with the same mass reveals that the gravitational acceleration for the disk model is weaker by a factor of $1/\sqrt{2}$ 1~kpc above the plane from simple Newtonian gravity calculations.  Thus while the fact that M82's potential well is not spherically symmetric can have a fairly large effect, and it implies a higher Eddington ratio, the effect is not large enough to affect our overall conclusions, since our fiducial estimate of $\Gamma$ is $\sim 0.02$.

The radiation pressure force on the grains depends largely on two things: their scattering and absorption properties, and the flux incident on them. The biggest determinant of the first is our choice of grain size distribution.  It is known that the MRN dust model used here does not reproduce the extinction curve towards, e.g., molecular clouds even in the Milky Way (see, for example, Weingartner \& Draine\ 2001, Cardelli et al.\ 1989), or the extinction present in starbursts (Calzetti et al.\ 1994); M82 dust could have a different size distribution and composition as well. However, by the analysis in Section~\ref{sec:results}, this likely would not affect the answer by more than a factor of $2-3$.

Our calculation of the flux incident on the grains is itself subject to several sources of uncertainty.  As seen in Equation~\ref{eq:Lstar}, it depends on the density of dust grains and their optical properties ($L_{\lambda}\propto (\alpha_{\lambda}\rho\kappa_{\rm s,\,\lambda})^{-1}$).  There are also three sources of reddening and extinction that go into calculating the flux:  material in the disk of M82 and the starburst itself, extinction in the wind, and material in our own galaxy.  We correct the last using the reddening corrections from Wyder et al.\ (2007), as described in Section~\ref{sec:results}.  We assumed that the wind was optically thin; we showed that this was true for the portion of the wind we have dust density data for in Section~\ref{sec:param}.  However, our results indicate that the starburst itself is optically thick in the UV ($\tau_{\rm FUV} \sim1.6 - 4$).  This large uncertainty in $\tau_{\rm FUV}$ introduces a similarly large uncertainty in $\Gamma$.

We correct for this reddening using the Calzetti et al.\ (1994) extinction law, as described in Section~\ref{sec:results}.  This correction increases $\Gamma$ by a factor of up to $\sim 1.5$.  However, the Calzetti et al. extinction law assumes the dust is in a foreground screen, while the results of F\"{o}rster Schreiber et al.\ (2001) indicate that modeling the dust and stars as intermixed is a better description of the M82 starburst.  In a series of comparisons between the shape of the emergent SED in the case of a foreground screen ($e^{-\tau}$) and a mixed disribution of sources and absorbers ($[1-e^{-\tau}]/\tau$), we find that the Eddington ratio might change by a factor of $\sim1.5-2$ at fixed $L_{\star,\,\rm UV}(z)$.

For the source geometry, we model the starburst as a point source in our order of magnitude calculations and our Monte Carlo simulations.  Past $\sim$1~kpc from the plane, this is a $\sim$5\% effect at most, so the uncertainty introduced in $\alpha_\lambda$ is small compared to our other uncertainties.  We also modeled the outer disk of M82 using STARBURST99 spectra as a 0.5~Gyr-old population of $4\times10^9\:{\rm M_{\odot}}$ (Mayya et al.\ 2006).  Even assuming no reddening or extinction between the starburst and the superwind, the outer disk in this model provides $\sim$1\% of the flux illuminating the wind, and therefore does not contribute significantly to the Eddington ratio.

We also assume that the dust density is constant along the line of sight for a given $z$.  While this cannot be strictly true, this assumption does not have a large effect on our calculations.  Although Equation~\ref{eq:Edd} does contain $\rho$ through $L_{*,\lambda}$ and Equation~\ref{eq:Lstar}, from Equation~\ref{eq:rad} we see that the observed flux of scattered light depends on the observed dust column, which is taken from Roussel et al.\ (2010).  Thus, the main factor that will change given $\rho (s)$ is $\alpha$.  We do not expect this change in $\alpha$ to heavily affect our results.

In addition to the extraction aperture through the GALEX data we describe in this paper (Figure~\ref{fig:M82im}), we also examined two other extraction apertures through the wind, one centered a few tens of arcseconds northeast of our first one, and another centered a few tens of arcseconds southwest.  We found that there was no difference in the overall conclusions; neither the dust densities nor the UV fluxes varied by more than a factor of two.  Only the detailed shapes of the distributions changed.

\section{Discussion and Conclusions} \label{sec:discussion}

Our results indicate that even though the escaping UV luminosity is high compared to previous observations (Hoopes et al.\ 2005), the overall escape fraction of UV light is only $\sim 1 - 15\%$.  Given that the dust opacity continues to rise towards the Lyman edge (Draine\ 2003b), it follows that there must be a low escape fraction of ionizing photons, likely less than $\sim 10\%$.  This is consistent with results obtained by several other groups (see, e.g., Heckman et al.\ 2001, Bergvall et al.\ 2006, Grimes et al.\ 2009, Heckman et al.\ 2011, Leitherer et al.\ 1995).  Heckman et al.\ (2001) found that galactic winds do not necessarily clear paths through the ISM for ionizing radiation to escape. It does not appear that M82's superwind has cleared such a path.

In this work, we generally treated the dust and gas as separate, except when we calculated the final Eddington ratio using Equation~\ref{eq:Edd}.  Ignoring magnetic fields and grain charging, which enhance dust-gas coupling, the scale over which grains of size $a$ are hydrodynamically coupled to the gas is (Draine \& Salpeter\ 1979, Murray et al.\ 2005)
\begin{equation}
\lambda_{\rm M} \simeq 10a_{0.1}\rho_{3}n_1^{-1}\:\mbox{pc}
\end{equation}
where $a_{0.1}=a/(0.1\:\mu{\rm m})$, $n_1=n/{\rm cm^{-3}}$ is the gas number density, and $\rho_3=\rho/3\,{\rm g\,\,cm^{-3}}$ is the individual grain density.  As shown by Figure~\ref{fig:dust}, typical gas densities in the wind are $0.1-0.01\:{\rm cm^{-3}}$, assuming a gas-to-dust ratio of 100, our model geometries for the dust density, and the Roussel et al.\ (2010) data.  Thus, $\lambda_{\rm M}$ ranges from $\sim 0.1-1\:{\rm kpc}$ for $0.1\:\mu{\rm m}$ grains, and $\sim 10-100\:{\rm pc}$ for smaller grains.  This means that the dust and gas are likely hydrodynamically coupled in the wind on large scales, and that the wind is sub-Eddington ($\Gamma \approx 0.01-0.06$ for $z\gtrsim 2$~kpc).

Although our work indicates that the M82 superwind is sub-Eddington on large scales, there is still the question of whether it is super-Eddington on small scales ($z\sim0.25-2$\,kpc).  If an MRN dust model and the rest of our assumptions (see Section~\ref{sec:results}) are valid inside the region where we have dust density data from Roussel et al.\ (2010), then we can extend the Eddington ratio calculation inwards, given a determination of the enclosed dynamical mass.  The rotation curve of GMT12 implies $M_\star(r\lesssim250\:{\rm pc})\simeq2-6\times 10^8$\,M$_{\odot}$, but on these small scales their calculation is affected by the stellar bar in M82.  F\"{o}rster Schreiber et al.\ (2001) estimate $M_\star(r\lesssim250\:{\rm pc})\simeq8\pm 2\times 10^8$\,M$_{\odot}$ from molecular and ionized gas tracers, similar, but somewhat higher than a naive application of GMT12.  Given these uncertainties, we estimate that for $z\approx250\:{\rm pc}$
\begin{equation}
\Gamma\sim0.1\,
\left(\frac{L_{\rm \star,\,UV}}{2\times10^{42}\,{\rm ergs\,\,s^{-1}}}\right)
\left(\frac{8\times10^8\,{\rm M_\odot}}{M_\star(r\lesssim250\,{\rm pc})}\right),
\label{gammain}
\end{equation}
where there are factor of $\sim2-3$ uncertainties in both $L_{\rm \star,\,UV}$ and $M_\star(r\lesssim250\,{\rm pc})$. The dashed curve in Figure \ref{fig:Edd} shows an explicit calculation of $\Gamma$ on small scales assuming the GMT12 mass estimate and $L_{\rm \star,\,UV}=2\times10^{42}$\,ergs s$^{-1}$.  Note that using the value of  $L_{\rm \star,\,UV}\simeq6\times10^{42}$\,ergs s$^{-1}$ (near the maximum in Figure \ref{fig:Lstar}) increases the dashed curve in Figure \ref{fig:Edd} by a factor of 3 at equivalent $\tau_{\rm UV}$.  Thus, this comparison indicates that the M82 superwind is currently sub-Eddington on $z\simeq0.25-0.5$\,kpc scales.

Another interesting possibility is that the superwind was super-Eddington in the past when the starburst was younger and brighter.  The models of F\"{o}rster Schreiber et al.\ (2003) indicate that the luminosity of the starburst was higher by about a factor of four roughly 6~Myr ago, for a bolometric luminosity of $2.6\times10^{11}\:{\rm L_{\odot}}$.  Assuming a similar dust and gas distribution to today's results in $\Gamma\sim 0.05 - 0.3$ on large scales and $\Gamma\sim 0.2-1$ on smaller scales.  However, if the gas surface density within the starburst was higher at that earlier time, this would lower the flux-mean opacity in the outflow, decreasing $\Gamma$ throughout the wind.

Finally, we consider the current Eddington limit for the dusty gas within the starburst itself. In this case, the appropriate Eddington limit is the single-scattering limit discussed in Murray et al.\ (2005), Thompson et al.~(2005), and Andrews \& Thompson (2011).  If all of the radiation is absorbed or scattered once, then the single-scattering Eddington flux is $F_{\rm Edd}\sim 2\pi G \Sigma_\star \Sigma_g c$ for a thin disk, where $\Sigma_\star$ and $\Sigma_g$ are the stellar and gas surface densities.  Scaling to parameters appropriate to M82's nucleus, $F_{\rm Edd}\sim4\times10^{11}\,\,{\rm L_\odot\,\,kpc^{-2}}\,(f_g/0.1)(M_\star/8\times10^8\,{\rm M_\odot})^2(250\,{\rm pc}/R)^4$,\footnote{Note that our use of $f_g=0.1$ is prima facie inconsistent with $\tau_{\rm UV}\simeq3$ for the obscuring screen used to model the emergent SED in Section \ref{sec:results}, since this UV optical depth would imply $\Sigma_g\sim{\rm few}\times10^{-3}$\,cm$^2$ g$^{-1}$ of gas.  However, as discussed in Section \ref{sec:error} a mixed medium of sources and obscuration is more realistic, and implies $\tau_{\rm UV}\simeq40\pm30$, which is consistent with $f_g=0.1$.} where $f_g=\Sigma_g/\Sigma_\star$ is the gas fraction, whereas the observed FIR flux of M82 is $F_{\rm FIR}\simeq3\times10^{11}\,{\rm L_\odot\,\,kpc^{-2}}(L_{\rm FIR}/6\times10^{10}\,\,L_\odot)(250\,{\rm pc}/R)^2$.  Within the uncertainties, it remains plausible that M82's nuclear starbust region currently exceeds the single-scattering Eddington limit, and may have significantly exceeded it 6~Myr ago, but clearly further work is required.  Lastly, we note that many of the super star clusters in M82 exceed the single-scattering Eddington limit (Krumholz \& Matzner 2009; Murray et al.~2010; Murray et al.~2011), and may thus be in part responsible for injecting gas into the hot outflow, or directly accelerating the gas to high velocities (Murray et al.~2011; but, see Krumholz \& Thompson 2013).

\section{Summary} \label{sec:summary}

We estimate the Eddington ratio for the dusty gas along the minor axis in the M82 superwind. We constrain the dust density distribution using Herschel SPIRE data and the dust mass map provided by Roussel et al.\ (2010), and we use Monte Carlo simulations to verify the applicability of the MRN dust distribution and both of our model geometries in our calculations by comparing the simulation results with archival GALEX images of M82.  We find that the combined UV luminosity in the GALEX bands that illuminates the superwind of M82 is $2-12$ times higher than that found by Hoopes et al.\ (2005) from our line of sight.

We find that the dusty gas in the wind is highly sub-Eddington ($\Gamma\sim 0.01 - 0.06$) on vertical scales along the minor axis $\gtrsim2$\,kpc (Figure~\ref{fig:Edd}). On smaller scales, $\Gamma$ is more uncertain, but increases because of the decrease in the enclosed mass. We estimate $\Gamma\sim0.05-0.3$ at $z\sim250$\,kpc, with large uncertainties, depending on the dynamical mass of the central region, the distribution of absorbing material along the minor axis, and the true UV luminosity that escapes the central starburst region (see Equation~\ref{gammain}).  Note that if the dust were hydrodynamically uncoupled from the gas, which appears unlikely, the grains would be super-Eddington throughout the wind.  In addition, the starburst was brighter by approximately a factor of four $\sim6$\,Myr ago (F\"{o}rster Schreiber et al.\ 2003), which in principle could to lead to $\Gamma$ as high as $\sim0.2-1$ at $z\sim250$\,kpc.     Finally, within the starburst itself, we find that M82 appears to be close to the single-scattering Eddington limit, implying that a significant fraction of the current ISM could be ejected by radiation pressure.  

Overall, our results provide quantitative constraints on theoretical models of superwinds launched wholly or in part by radiation pressure on dust in dwarf starbursts like M82 (e.g., Murray et al.~2011; Hopkins et al.~2012).

There are a number of directions for future work.  Perhaps most importantly, one could simultaneously model the scattering in a number of wavebands from the FUV through the NIR, using the full frequency-dependent dust scattering phase function.  By iterating with the observed surface brightness profiles in each band, a self-consistent model might be developed.  Information could also be included on the MIR PAH surface brightness as a function of height.  Such a study would constrain the dust geometry and grain size distribution more fully and perhaps allow for novel studies of the starburst population in reflection off the wind.

\begin{acknowledgements}

We thank Chris Kochanek for useful conversations and Mark Krumholz, Eliot Quataert, and Norm Murray for comments.  TAT thanks Brian Lacki for useful discussions and collaboration at an early stage in formulating this paper.  We thank Helene Roussel for providing us with her dust map of M82.  This work supported in part by NASA grant NNX10AD01G.

\end{acknowledgements}

\end{document}